\documentclass[article,superscriptaddress,twocolumn,amsmath,amssymb,floatfix]{revtex4}
\usepackage{graphicx,psfig}
\usepackage{dcolumn}
\usepackage{bm}
\begin{document}

\title{Maximum likelihood: extracting unbiased information from complex networks}
\author{Diego Garlaschelli}
\affiliation{Dipartimento di Fisica, Universit\`a di Siena, Via Roma 56, 53100 Siena ITALY.}
\author{Maria I. Loffredo}
\affiliation{Dipartimento di Scienze Matematiche ed Informatiche, Universit\`a di Siena, Pian dei Mantellini 44, 53100 Siena ITALY.}
\begin{abstract}
The choice of free parameters in network models is subjective, since it depends on what topological properties are being monitored. However, we show that the Maximum Likelihood (ML) principle indicates a unique, statistically rigorous parameter choice, associated to a well defined topological feature. 
We then find that, if the ML condition is incompatible with the built-in parameter choice, network models turn out to be intrinsically ill-defined or biased.
To overcome this problem, we construct a class of safely unbiased models.
We also propose an extension of these results that leads to the fascinating possibility to extract, only from topological data, the `hidden variables' underlying network organization, making them `no more hidden'. We test our method on the World Trade Web data, where we recover the empirical Gross Domestic Product using only topological information.
\end{abstract}
\maketitle
In complex network theory, graph models are systematically used either as null hypotheses against which real--world networks are analysed, or as testbeds for the validation of network formation mechanisms \cite{guidosbook}. 
Until now there has been no rigorous scheme to define network models. 
However, here we use the Maximum Likelihood (ML) principle to show that undesired statistical biases naturally arise in graph models, which in most cases turn out to be ill--defined. We then show that the ML approach constructively indicates a correct definition of unbiased models. Remarkably, it also allows to extract hidden information from real networks, with intriguing consequences for the understanding of network formation. 

The framework that we introduce here allows to solve three related, increasingly complicated problems. 
First, we discuss the correct choice of free parameters. 
Model parameters are fixed in such a way that the expected values (i.e. ensemble averages over many realizations) of some `reference' topological property match the empirically observed ones. 
But since there are virtually as many properties as we want to monitor in a network, and surely many more than the number of model parameters, it is important to ask if the choice of the reference properties is arbitrary or if a rigorous criterion exists. We find that the ML method provides us with a unique, statistically correct parameter choice.
Second, we note that the above ML choice may conflict with the structure of the model itself, if the latter is defined in such a way that the expected value of some property, which is not the correct one, matches the corresponding empirical one. We find that the ML method identifies such intrinsically ill--defined models, and can also be used to define safe, unbiased ones.
The third, and perhaps most fascinating, aspect regards the  extraction of information from a real network. Many models are defined in terms of additional `hidden variables' \cite{fitness,soderberg,pastor,chunglu} associated to vertices. The ultimate aim of these models is to identify the hidden variables with empirically observable quantities, so that the model will provide a mechanism of network formation driven by these quantities. While for a few networks this identification has been carried out successfully \cite{mywtw,myshares}, in most cases the hidden variables are assigned \emph{ad hoc}. However, since in this case the hidden variables play essentially the role of free parameters, one is led again to the original problem: if a non--arbitrary parameter choice exists, we can infer the hidden variables from real data. As a profound and exciting consequence, the quantities underlying network organization are `no more hidden'.

In order to illustrate how the ML method solves this three--fold problem successfully, we use equilibrium graph ensembles as an example. 
All network models depend on a set of parameters that we collectively denote by the vector $\vec{\theta}$. 
Let $P(G|\vec{\theta})$ be the conditional probability of occurrence of a  graph $G$ in the ensemble spanned by the model.
For a given topological property $\pi(G)$ displayed by a graph $G$, the expected value $\langle \pi\rangle_{\vec{\theta}}$ reads
\begin{equation}
\langle \pi\rangle_{\vec{\theta}}\equiv\sum_G \pi(G)P(G|\vec{\theta})
\label{eq_expected}
\end{equation}
In order to reproduce a real--world network $A$, one usually chooses some reference properties $\{\pi_i\}_i$ and then sets $\vec{\theta}$ to the `matching value' $\vec{\theta}_{M}$ such that 
\begin{equation}
\langle \pi_i\rangle_{\vec{\theta}_{M}}=\pi_i(A)\quad\forall i
\label{eq_match}
\end{equation}
Our first problem is: is this method statistically rigorous? And what properties have to be chosen anyway?
A simple example is when a real undirected network $A$ with $N$ vertices and $L$ undirected links is compared with a random graph where the only parameter is the connection probability $\theta=p$. The common choice for $p$ is such that the expected number of links $\langle L\rangle_{p}=p N(N-1)/2$ equals the empirical value $L$, which yields $p_{M}=2L/N(N-1)$. But one could alternatively choose $p$ in such a way that the expected value $\langle C\rangle$ of the clustering coefficient matches the empirical value $C$, resulting in the different choice $p_M=C$. Similarly, one could choose any other reference property $\pi$, and end up with different values of $p$. Therefore, in principle the optimal choice of $p$ is undetermined, due to the arbitrariness of the reference property.

However, we now show that the ML approach indicates a unique, statistically correct parameter choice. Consider a random variable $v$ whose probability distribution $f(v|\theta)$ depends on a parameter $\theta$. 
For a physically realized outcome $v=v'$, $f(v'|\theta)$ represents the \emph{likelihood} that $v'$ is generated by the parameter choice $\theta$.
Therefore, for fixed $v'$, the optimal choice for $\theta$ is the value $\theta^*$ maximizing $f(v'|\theta)$ or equivalently $\lambda(\theta)\equiv\log f(v'|\theta)$. The ML approach avoids the drawbacks of other fitting methods, such as the subjective choice of fitting curves and of the region where the fit is performed. This is particularly important for networks, often characterized by broad distributions that may look like power laws with a certain exponent (subject to statistical error) in some region, but that may be more closely reproduced by another exponent or even by different curves as the fitting region is changed. By contrast, the ML approach always yields a unique and rigorous parameter value. Examples of recent applications of the ML principle to networks can be found in \cite{ml,newman_likelihood}.
In our problem,  the likelihood that a real network $A$ is generated by the parameter choice $\vec{\theta}$ is  
\begin{equation}
\lambda(\vec{\theta})\equiv \log P(A|\vec{\theta})
\label{eq_likelihood}
\end{equation}
and the ML condition for the optimal choice $\vec{\theta}^*$ is
\begin{equation}
\vec{\nabla}\lambda(\vec{\theta}^*)=\left[\frac{\partial \lambda(\vec{\theta})}{\partial \vec{\theta}}\right]_{\vec{\theta}=\vec{\theta}^*}=\vec{0}
\label{eq_derive}
\end{equation}
This gives a unique solution to our first problem. 
For instance, in the random graph model we have
\begin{equation}
P(A|p)=p^L(1-p)^{N(N-1)/2-L}
\end{equation}
Writing the likelihood function $\lambda(p)=\log P(A|p)$ and looking for the ML value $p^*$ such that $\lambda'(p^*)=0$ yields 
\begin{equation}
p^*=\frac{2L}{N(N-1)}
\end{equation}
Therefore we find that the ML value for $p$ is the one we obtain by requiring  $\langle L\rangle=L$. In general, different reference quantities (for instance the clustering coefficient) would not yield the statistically correct ML value. 

For the random graph model the above correct choice is also the most frequently used. However, more complicated models may be intrinsically ill--defined, as there may be no possibility to match expected and observed values of the desired reference properties without violating the ML condition. 
This is the second problem we anticipated. To illustrate it, it is enough to consider a slightly more general class of models, obtained when the links between all pairs of vertices $i,j$ are drawn with different and independent probabilities $p_{ij}(\vec{\theta})$ \cite{fitness,soderberg,pastor,chunglu}. Now
\begin{equation}
P(A|\vec{\theta})=\prod_{i<j}p_{ij}(\vec{\theta})^{a_{ij}}[1-p_{ij}(\vec{\theta})]^{1-a_{ij}}
\label{eq_paij}
\end{equation}
where the product runs over vertex pairs $(i,j)$, and $a_{ij}=1$ if $i$ and $j$ are connected in graph $A$, and $a_{ij}=0$ otherwise. Then eq.(\ref{eq_likelihood}) becomes
\begin{equation}
\lambda(\vec{\theta})=\sum_{i<j}a_{ij}\log\frac{p_{ij}(\vec{\theta})}{1-p_{ij}(\vec{\theta})}+\sum_{i<j}\log[1-p_{ij}(\vec{\theta})]
\label{eq_likelihoodij}
\end{equation}
For instance, in hidden variable models \cite{fitness,soderberg,pastor} $p_{ij}$ is a function of a control parameter $\theta\equiv z$ and of some quantities $x_i$, $x_j$ that we assume fixed for the moment. 
As a first example, consider the popular bilinear choice \cite{fitness,soderberg,pastor,chunglu}
\begin{equation}
p_{ij}(z)=zx_ix_j
\label{eq_CL}
\end{equation}
Writing $\lambda(z)=\log P(A|z)$ as in eq.(\ref{eq_likelihoodij}) and deriving yields
\begin{equation}
\lambda'(z^*)=\sum_{i<j}\left[\frac{a_{ij}}{z^*}-\frac{(1-a_{ij})x_ix_j}{1-z^*x_ix_j}\right]=0
\end{equation}
Since $\sum_{i<j}a_{ij}=L$, the condition for $z^*$ becomes
\begin{equation}
L=\sum_{i<j}(1-a_{ij})\frac{z^*x_ix_j}{1-z^*x_ix_j}
\label{eq_CL_ML}
\end{equation}
This shows that if we set $z=z^*$, then $L$ is in general different from the expected value $\langle L\rangle_{z^*}=\sum_{i<j}p_{ij}(z^*)=\sum_{i<j}z^*x_ix_j$. 
This means that if we want the ML condition to be fulfilled, we cannot tune the expected number of links to the real one! Viceversa, if we want the expected number of links to match the empirical one, we have to set $z$ to a value different from the statistically correct $z^*$ one. 
The problem is particularly evident since, setting $x_i\equiv \langle k_i\rangle/\sqrt{\langle L\rangle}$, eq.(\ref{eq_CL}) can be rewritten as $p_{ij}=\langle k_i\rangle\langle k_j\rangle/(2\langle L\rangle)$ \cite{chunglu}.
So, in order to reproduce a network with $L$ links we should paradoxically set the built--in parameter $\langle L\rangle=(2z)^{-1}$ to a ML value which is different from $L$. In analogy with the related problem of biased estimators in statistics, we shall define a \emph{biased model} any such model where the use of eq.(\ref{eq_match}) to match expected and observed properties violates the ML condition. 
As a second example, consider the model \cite{mywtw,newman_origin,parknewman} 
\begin{equation}
p_{ij}(z)=\frac{zx_ix_j}{1+zx_ix_j}
\label{eq_hiddenOK}
\end{equation}
Writing $\lambda(z)$ and setting $\lambda'(z^*)=0$ now yields
\begin{equation}
L=\sum_{i<j}\frac{z^*x_ix_j}{1+z^*x_ix_j}
\label{eq_Lhidden}
\end{equation}
which now coincides with $\langle L\rangle_{z^*}=\sum_{i<j}p_{ij}(z^*)$, 
showing that this model is unbiased: the ML condition (\ref{eq_derive}) and the requirement $\langle L\rangle=L$ are equivalent. In a previous paper \cite{mywtw}, we showed that this model reproduces the properties of the World Trade Web (WTW) once $x_i$ is set equal to the Gross Domestic Product (GDP) of the country represented by vertex $i$. The parameter $z$ was chosen as in eq.(\ref{eq_Lhidden}) \cite{mywtw}, and now we find that this is the correct criterion. We shall again consider the WTW later on.

The above examples show that while some models are unbiased, others are `prohibited' by the ML principle. The problem of bias potentially underlies all network models, and is therefore of great importance. Is there a way to identify the class of safe, unbiased models? We now show that one large class of unbiased models can be constructively defined, namely the exponential random graphs traditionally used by sociologists \cite{holland,WF} and more recently considered by physicists \cite{parknewman,burda,berg,holyst}. 
If $\{\pi_i\}_i$ is a set of topological properties, an exponential model is defined by the probability
\begin{equation}
P(G|\vec{\theta})=e^{-H(G|\vec{\theta})}/Z(\vec{\theta})
\label{eq_exp}
\end{equation}
where $H(G|\vec{\theta})\equiv\sum_i \pi_i(G)\theta_i$ is the graph Hamiltonian and $Z(\vec{\theta})\equiv \sum_G \exp [-H(G|\vec{\theta})]$ is the partition function \cite{parknewman,burda,berg,holyst}. In the standard approach, one chooses the matching value $\vec{\theta}_M$ fitting the properties of a real network. In order to check whether this violates the ML principle, we need to look for the value $\vec{\theta}^*$ maximizing the likelihood to obtain a network described by a given set $\{\pi_i\}_i$ of reference properties.
The likelihood function we have defined reads 
$\lambda(\vec{\theta})\equiv \log P(A|\vec{\theta})=-H(A|\vec{\theta})-\log Z(\vec{\theta})$
and eq.(\ref{eq_derive}) gives for $\vec{\theta}^*$ 
\begin{equation}
\left[\frac{\partial \lambda(\vec{\theta})}{\partial \theta_i}\right]_{\vec{\theta}=\vec{\theta}^*}\!\!\!=
\left[-\pi_i(A)-\frac{1}{Z(\vec{\theta})}\frac{\partial Z(\vec{\theta})}{\partial\theta_i}\right]_{\vec{\theta}=\vec{\theta}^*}\!\!\!=0
\end{equation}
whose solution yields the ML condition 
\begin{equation}
\pi_i(A)=\sum_G \pi_i(G)e^{-H(G|\vec{\theta}^*)}/Z(\vec{\theta}^*)=\langle \pi_i\rangle_{\vec{\theta}^*}\quad\forall i
\label{eq_system}
\end{equation}
which is equivalent to eq.(\ref{eq_match}): remarkably, $\vec{\theta}^*=\vec{\theta}_{M}$ and the model is unbiased. 
We have thus proved a remarkable result: any model of the form in eq.(\ref{eq_exp}) is unbiased under the ML principle, if and only if all the properties $\{\pi_i\}_i$ included in $H$ are simultaneously chosen as the reference ones used to tune the parameters $\vec{\theta}$. The statistically correct values $\vec{\theta}^*$ of the latter are the solution of the system of (in general coupled) equations (\ref{eq_system}). There are as many such equations as the number of free parameters.
This gives us the following recipe: if we are defining a model whose predictions will be matched to a set of properties $\{\pi_i(A)\}_i$ observed in a real--world network $A$, we should decide from the beginning what these reference properties are, include them in $H(G|\vec{\theta})$ and define $P(G|\vec{\theta})$ as in eq.(\ref{eq_exp}). In this way we are sure to obtain an unbiased model. The random graph is a trivial special case where $\pi(A)=L$ and $H(G|\theta)=\theta L$ with $p\equiv (1+e^\theta)^{-1}$ \cite{parknewman}, and this is the reason why it is unbiased, if $L$ is chosen as reference. 
The hidden--variable model defined by eq.(\ref{eq_hiddenOK}) is another special case where $\pi_i(A)=k_i$ and $H(G|\vec{\theta})=\sum_i \theta_i k_i$ with $x_i\equiv e^{-\theta_i}$ \cite{parknewman}, and so it is unbiased too. By contrast, eq.(\ref{eq_CL}) cannot be traced back to eq.(\ref{eq_exp}), and the model is biased.
Once the general procedure is set out, one can look for other special cases. 
The field of research on exponential random graphs is currently very active\cite{parknewman,burda,berg,holyst,mygrandcanonical,holyst_clustering}, and models including correlations and higher--order properties are being studied, for instance to explore graphs with nontrivial reciprocity \cite{mygrandcanonical} and clustering \cite{holyst_clustering}. For each of these models, our result (\ref{eq_system}) directly yields the unbiased parameter choice in terms of the associated reference properties.

We can now address the third problem.
In the cases considered so far we assumed that the values of the hidden variables $\{x_i\}_i$ were pre--assigned to the vertices. This occurs when we have a candidate quantity to identify with the hidden variable \cite{mywtw,myshares}. However we can reverse the point of view and extend the ML approach so that, without any prior information, the hidden variables are included in $\vec{\theta}$ and treated as free parameters themselves, to be tuned to their ML values $\{x^*_i\}_i$. In this way, hidden variables will be no longer `hidden', since they can be extracted from topological data. This is an exciting possibility that can be applied to any real network. Moreover, this extension of the parameter space also allows us to match $N$ additional properties besides the overall number of links. However, the unbiased choice of these properties must be dictated by the ML principle. 

For instance, 
let us look back at the model defined in eq.(\ref{eq_hiddenOK}), now considering $x_i$ and $x_j$ not as fixed quantities, but as free parameters exactly as $z$, to be included in $\vec{\theta}$. 
Deriving $\lambda(\vec{\theta})=\lambda(z,x_1,\dots, x_N)$ with respect to $z$ gives again eq.(\ref{eq_Lhidden}) with $x_i$ replaced by $x_i^*$, and deriving with respect to $x_i$ yields the $N$ additional equations
\begin{equation}
k_i=\sum_{j\ne i}\frac{z^*x^*_ix^*_j}{1+z^*x^*_ix^*_j}\qquad i=1,\dots,N
\label{eq_Khidden}
\end{equation}
Therefore we find that the $N$ correct reference properties for this model are the degrees: $\langle k_i\rangle_{\vec{\theta}^*}=\sum_{j\ne i} p_{ij}(\vec{\theta}^*)=k_i$. 
This is not true in general: the model (\ref{eq_CL}) would imply different reference properties such that $\langle k_i\rangle\ne k_i$, so that choosing the degrees as the properties to match would bias the parameter choice. 
Again, this difference arises because eq.(\ref{eq_Khidden}) corresponds to eq.(\ref{eq_system}) for  the exponential model  $H(G|\vec{\theta})=\sum_i \theta_i k_i$ \cite{parknewman}, while the model in eq.(\ref{eq_CL}) cannot be put in an exponential form.
We stress that, although eq.(\ref{eq_Khidden}) is formally identical to the familiar expression yielding $\langle k_i\rangle$ as a function of $\{x_i\}_i$ if the latter are fixed \cite{parknewman}, its meaning here is completely reversed: the degrees $k_i$ are fixed by observation and the unknown hidden variables are inferred from them through the ML condition. 
This is our key result. 
Note that, although determining the $x_i^*$'s requires to solve the $N+1$ coupled equations (\ref{eq_Lhidden}) and (\ref{eq_Khidden}), the number of independent expressions is much smaller since: i) eqs.(\ref{eq_Khidden}) automatically imply eq.(\ref{eq_Lhidden}), so we can reabsorbe $z^*$ in a redefinition of $x^*_i$ and discard eq.(\ref{eq_Lhidden}); ii) all vertices with the same degree $k$ obey equivalent equations and hence are associated to the same value $x^*_k$. So eqs.(\ref{eq_Khidden}) reduce to 
\begin{equation}
k=\sum_{k'}P(k')\frac{x^*_k x^*_{k'}}{1+x^*_k x^*_{k'}}-\frac{(x^*_k)^2}{1+(x^*_k)^2}
\label{eq_k}
\end{equation}
where $P(k)$ is the number of vertices with degree $k$, the last term removes the self--contribution of a vertex to its own degree, and $k$ and $k'$ take only their empirical values. Hence the number of nonequivalent equations equals the number of distinct degrees that are actually observed, which is always much less than $N$. 
\begin{figure}	
\includegraphics[width=.37\textwidth]{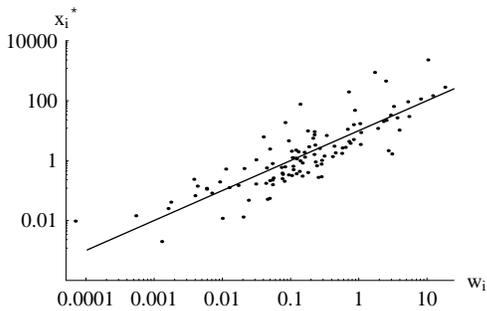}
\caption[]{\small ML hidden variables ($x^*_i$) versus GDP rescaled to the mean ($w_i$) for the WTW (year 2000), and linear fit.\label{fig}}
\end{figure}

We can test our method on the WTW data, since from the aforementioned previous study we know that the GDP of each country plays the role of the hidden variable $x_i$, and that the real WTW is well reproduced by eq.(\ref{eq_hiddenOK}) \cite{mywtw}. We can first use eq.(\ref{eq_k}) to find the values $\{x^*_i\}_i$ by exploiting only topological data (the degrees $\{k_i\}_i$), and then compare these values with the empirical GDP of each country $i$ (which is independent of topological data), rescaled to its mean to factor out physical units. As shown in fig.\ref{fig}, the two variables ideed display a linear trend over several orders of magnitude. Therefore our method identifies the GDP as the hidden variable successfully. 
Clearly, our approach can be used to uncover hidden factors from other real--world networks, such as biological and social webs. 
An example is that of food web \cite{niche} models, where it is assumed that predation probabilities depend on hypothetical \emph{niche} values $n_i$ associated to each species. Our formalism allows to extract niche values directly from empirical food webs, and not from \emph{ad hoc} statistical distributions \cite{niche}. Another interesting application is to gene regulatory networks, where the length of regulatory sequences and promoter regions have been shown to determine the connection probability $p_{ij}$\cite{duygu}. 
Similarly, our approach allows to extract the vertex--specific quantities (such as \emph{expansiveness}, \emph{actractiveness} or \emph{mobility}--related parameters) that are commonly assumed to determine the topology and community structure of social networks \cite{holland,WF,mobility}.
In all these cases, the hypotheses can be tested against real data by plugging any particular form of $p_{ij}=p(x_i,x_j)$ into eq.(\ref{eq_likelihoodij}) and looking for the values $\{x^*_i\}_i$ 
that solve eq.(\ref{eq_derive}), i.e.
\begin{equation}
\sum_{j\ne i}
\frac{a_{ij}-p(x^*_i,x^*_j)}{p(x^*_i,x^*_j)[1-p(x^*_i,x^*_j)]}
\left[\frac{\partial p(x_i,x_j)}{\partial x_i}\right]_{\vec{x}=\vec{x}*}
\!\!\!=0\quad\forall i
\label{eq_general}
\end{equation}
Note that for eq.(\ref{eq_hiddenOK}) one correctly recovers eq.(\ref{eq_Khidden}).
Once obtained, the values $\{x^*_i\}_i$ can be compared with the (totally independent) empirical ones to check for significant correlations, as we have done for the GDP data. Clearly, an important open problem to address in the future is understanding the conditions under which eq.(\ref{eq_general}), and similarly eq.(\ref{eq_k}) for a generic $P(k)$, can be solved.

We have shown that the ML principle indicates the statistically correct parameter values of network models, making the choice of reference properties no longer arbitrary. It also identifies undesired biases in graph models, and allows to overcome them constructively. Most importantly, it provides an elegant way to extract information from a network by uncovering the underlying hidden variables. This possibility, that we have empirically tested in the case of the World Trade Web, opens to a variety of applications in economics, biology, and social science.

After submission of this article, we got aware of later studies based on a similar idea \cite{newman_likelihood,ramascomungan}.

\end{document}